\documentclass[runningheads]{llncs}
\usepackage{amsmath,amssymb}
\usepackage{comment}
\usepackage[utf8]{inputenc}
\usepackage{hyperref}
\usepackage{graphicx}
\usepackage[font=small]{caption}
\usepackage[caption=false]{subfig}
\usepackage{xcolor}
\usepackage{dsfont}
\usepackage{booktabs}
\usepackage{bm}
\usepackage[capitalise]{cleveref}

\begin{document}
	\title{Opinion Dynamics with Multi-Body Interactions\thanks{Parts of this work have been presented in~ \cite{neuhauser_multibody_2020}. 
    L. Neuhäuser was financially supported by the Hertie School. 
    M. Schaub was supported by the European Union’s Horizon 2020 research and innovation programme under the Marie Sklodowska-Curie grant agreement No 702410.}}
	%
	%
	\author{Leonie~Neuhäuser\inst{1,2}  \orcidID{0000-0002-0776-7031} \and
		Michael~T.~Schaub\inst{3}\orcidID{0000-0003-2426-6404} \and
		Andrew~Mellor\inst{2}\orcidID{0000-0002-1581-9671} \and
		Renaud~Lambiotte\inst{2}\orcidID{0000-0002-0583-4595}}
	\authorrunning{L. Neuhäuser et al.}
	%
	\institute{Hertie School, Berlin, Germany \and
	Mathematical Institute, University of Oxford, Oxford, UK \newline
		\email{\{neuhauser, mellor, renaud.lambiotte\}@maths.ox.ac.uk}\\ \and
		Dept. of Engineering Science, University of Oxford, Oxford, UK\\
		\email{michael.schaub@eng.ox.ac.uk}}
	\maketitle              
	\begin{abstract}
	We introduce and analyse a three-body consensus model (3CM) for non-linear consensus dynamics on hypergraphs. 
	Our model incorporates reinforcing group effects, which can cause shifts in the average state of the system even in if the underlying graph is complete (corresponding to a mean-field interaction), a phenomena that may be interpreted as a type of peer pressure. 
	We further demonstrate that for systems with two clustered groups, already a small asymmetry in our dynamics can lead to the opinion of one group becoming clearly dominant.
	We show that the nonlinearity in the model is the essential ingredient to make such group dynamics appear, and demonstrate how our system can otherwise be written as a linear, pairwise interaction system on a rescaled network.
		
	\end{abstract}

	\section{Introduction}
	Networks provide a powerful framework for the modelling of dynamical systems.
	Many networked dynamical systems can be described by a set of differential equations describing \emph{pairwise} interactions between the nodes:
	\begin{align}\label{eq:pairwise_dyn}
	\dot{x}_i =  \sum_j A_{ij} f(x_i,x_j) \qquad \text {for} \quad i \in \{1, \ldots, N\}.
	\end{align}
	where $x_i$ is the state of node $i$, $A\in\mathbb{R}^{N\times N}$ is the adjacency matrix of the underlying graph ($A_{ij} = 1$, if node $i$ connects to node $j$, and $0$ otherwise), and $f(x_i,x_j)$ is a function describing the interactions between nodes $i$ and $j$.
    Important examples of the above type of dynamics include diffusion \cite{masuda_random_2017} or oscillator dynamics \cite{arenas2008synchronization}.
    In particular, the dynamics of opinion formation have been considered in the context of dynamical processes on networks~\cite{Proskurnikov2017,Proskurnikov2018}, including opinion formation models such as the de Groot model \cite{DeGroot1974}, bounded-confidence models \cite{deffuant2000mixing} and threshold models \cite{watts2002simple}.
	
	However, it is increasingly realized that such pairwise interaction models may not be sufficient to describe a range of important phenomena, ranging from collaborations of authors \cite{patania_shape_2017} to neuronal activity \cite{giusti_clique_2015}.
	Accordingly, various models that focus on the importance of group interactions, i.e., situations when the basic unit of interaction involves more than two nodes have been proposed in the literature \cite{iacopini_simplicial_2019,arruda2019social,schaub_random_2018}.
	
	These \emph{multi-body} interaction models are particularly relevant for social dynamics, which have long been argued to be emergent phenomena that are not merely based on pairwise interactions between members of a community, but often include complex mechanisms of peer influence and reinforcement.  
	Such group dynamics, which may lead to `higher-order' dynamical effects, may indeed be essential to better understand phenomena such as hate communities, echo chambers and polarisation in society.

	\section{A Multi-body interaction model for non-linear consensus}
	Motivated by the above discussed scenarios, we here introduce a simple multi-body interaction model for opinion formation within social systems.
	As a first step towards studying the higher-order effects of multi-body dynamics, we concentrate on a three-body consensus model (3CM), in which the interactions between triplets of nodes are governed by the following differential equations: 
	\begin{align}\label{eq:dynamics}
	\dot{x}_{i}=\sum_{j , k=1}^N \mathcal A_{ijk} g^{\{j,k\}}_{i}(x_i,x_j,x_k) \quad \text {for } \; i \in \{1, \ldots, N\}.
	\end{align}
	Here $\mathcal A_{ijk}$ describes the adjacency tensor of node triplets $\{i,j,k\}$, where $\mathcal A_{ijk} = 1$ if the group of nodes interact and $\mathcal A_{ijk} = 0$ otherwise.
	We further model the group (multi-body) interaction function $g^{\!\{j,k\}}_{i}(x_i,x_j,x_k)$ as:
	\begin{align}\label{eqn:our_function}
	g^{\{j,k\}}_{i}(x_i,x_j,x_k)=s(\left\|x_j\!-\!x_k\right\|)\left[(x_j\!-\!x_i)+(x_k\!-\!x_i)\right].
	\end{align}
	For each triplet $\{i,j,k\}$, this function comprises (a) the joint influence of the node-pair $j,k$ on node $i$, modeled by the linear term $[(x_j-x_i)+(x_k-x_i)]$, which is (b) modulated by an influence function $s(\left\|x_j-x_k\right\|)$ of their state differences.
	In the following we assume $g^{\{j,k\}}_{i}$ is the same for each interacting node triplet, for the sake of simplicity.
	
	Note that if the modulation function $s(x)$ is monotonically decreasing, nodes $j$ and $k$ reinforce their influence on $i$ if they have similar states $x_j$, $x_k$, whereas the influence of nodes $j,k$ on node $i$ is diminished if their states are very different.
	This property is reminiscent of non-linear voter models for discrete dynamics \cite{lambiotte_dynamics_2008}, where voters change opinion with a probability depending non-linearly on the fraction of disagreeing neighbours. 
	
    In addition to the ability to describe a reinforcing dynamics, the functional form of our model has some further desirable symmetry properties.
    In particular, we remark that~\eqref{eq:dynamics} is invariant to translation and rotation of all node states.
    This is a desirably property for many opinion formation process, as it ensures that the opinion formation is only influenced by the relative position of the node states $x_i$ and independent of a specific global reference frame.
    This property can be shown by observing that any rotation is norm preserving, and thus $s(\|x_j-x_k\|)$  is rotational and translational invariant.
    Since the term $[(x_j-x_i)+(x_k-x_i)]$ is translation invariant and linear, any translation and rotation applied to all states will leave~\eqref{eq:dynamics} invariant.
    Note that this `quasi-linearity' of the interaction function $g^{\{j,k\}}_{i}(x_i,x_j,x_k)$ is in close correspondence to the necessary and sufficient conditions for translation and rotational invariance for pairwise interaction systems~\cite{vasile_sen_2015}.
    In the following we will restrict our scope to scalar states $x_i$. 
    In this case, the above described invariance simply implies an invariance under a change of signs or a global shift of all states.
		
	\section{Results}
	\subsection{Reduction to network model and higher-order effects}
	Interestingly, it can be shown that the above dynamics can be rewritten in terms of a (in general) \emph{time-varying and state dependent} weighted adjacency matrix $\mathfrak{W}(t,x,\mathcal A)$, whose entries describe the three-body influence on node $i$ exerted over the 'pairwise link' $(i,j)$:
	\begin{align}
	\label{eq:hello}
	(\mathfrak{W})_{ij}= \sum_{k}\mathcal A_{ijk}s(\left\|x_j-x_k\right\|)=\sum_{k \in \mathcal{I}_{ij}}s(\left\|x_j-x_k\right\|).
	\end{align}
	Here $\mathcal{I}_{ij}$ is the index-set of nodes that interact in a triplet with nodes $i,j$; and for simplicity we have written $\mathfrak{W}=\mathfrak{W}(t,x,\mathcal A)$, omitting the dependencies of $\mathfrak{W}$.
	Accordingly, we can rewrite the dynamics~(\ref{eq:dynamics}) as:
	\begin{align}
	\label{eq:ren}
	\dot{x}_i  = 2\sum_{j}\mathfrak{W}_{ij}(x_j-x_i)=: -2\sum_j L^\mathfrak{W}_{ij} x_j,
	\end{align}
	where we have defined the Laplacian $L^\mathfrak{W}$ of the 3CM model via  $L^\mathfrak{W}_{ij} = -\mathfrak{W}_{ij}$ for $i\neq j$, and $L^\mathfrak{W}_{ii}=\sum_j \mathfrak{W}_{ij}$.
	
	As discussed above, the entries of the weighted adjacency matrix $\mathfrak{W}$ (and thus of $L^\mathfrak{W}$) are in general time-varying, state and topology dependent for a general modulation function $s(t)$, and so the above rewriting does not imply that the system can be understood via pairwise interaction of the form (\ref{eq:pairwise_dyn}).
	There is one important exception, though. 
	If $s(x)$ is constant, the group interaction function $g$ is linear and the three-body dynamical system can therefore be rewritten as a rescaled pairwise dynamical system defined on a graph.
	The weighted adjacency matrix $\mathfrak{W}$ and corresponding graph Laplacian $L^\mathfrak{W}$ are then constant in time.
	In fact, in this case $L^\mathfrak{W}$ becomes the so-called motif Laplacian proposed by \cite{benson_higher_order_2016} for community detection in higher-order networks: the entries of $L^\mathfrak{W}$ count the nodes involved in interaction triplets (triangles on the corresponding graph).
	This emphasizes that multi-body dynamical effects beyond rescaled pairwise interactions can only appear for non-linear interaction functions, regardless of topology of the multi-body interactions encoded in $\mathcal A$.
	
	\subsection{Convergence to consensus and average-opinion dynamics}
	From our above rewriting~\eqref{eq:ren}, it is easy to see that a global consensus, in which $x_i(t) = x_j(t)$ for all $i,j$, is a fixed point of our model. 
	Using standard arguments, it can be shown that convergence to consensus is guaranteed as long as the scaling $s(\|x_j-x_k\|)$ is positive.  
	For generic initial conditions this is only the case if the modulation function $s(x)$ is positive definite. 
	We will therefore focus on this scenario in the following. 
	
	Despite the fact that in our model the 3-body interactions as undirected, i.e., the adjacency tensor $\mathcal{A}$ is completely symmetric in all pairs of indices, the average opinion $\bar{x} = (\sum_i x_i)/N$ is however not invariant over time, in general.
	To see this observe that
	\begin{align}
	\dot{\bar{x}}(t)= \frac{1}{N}\sum_{i=1}^N\dot{x}_i(t) & = -\frac{2}{N}\sum_{i,j=1}^N  L^\mathfrak{W}_{ij}x_j(t),
	\end{align}
	which is zero only (i) when there is global consensus, or (ii) when the Laplacian $L^\mathfrak{W}$ (interpreted as a graph Laplacian of a directed graph) corresponds to a balanced graph, i.e., the in-degree equals the out-degree for every node and thus  $\sum_i L^\mathfrak{W}_{ij}= \mathbf{0}$.
	While the former condition is dynamically trival, the latter condition will in general depend on a complex interplay between the states $x_i$, the structure of the node-triplet interaction and the form of the interaction function.
	One exception here is again the case in which the modulation function $s(x)$ is constant and the dynamics therefore becomes linear. 
	In this case $L^\mathfrak{W}$ will be a symmetric matrix and therefore trivially correspond to a balanced (undirected) graph.
	Note that if at any instance of time $t_0$ the Laplacian $L^\mathfrak{W}$ becomes balanced, the average opinion will be conserved for all $t\ge t_0$.
	
	\begin{figure}
		\centering
		\subfloat[$\lambda=-1$]{%
			\includegraphics[width=0.5\linewidth]{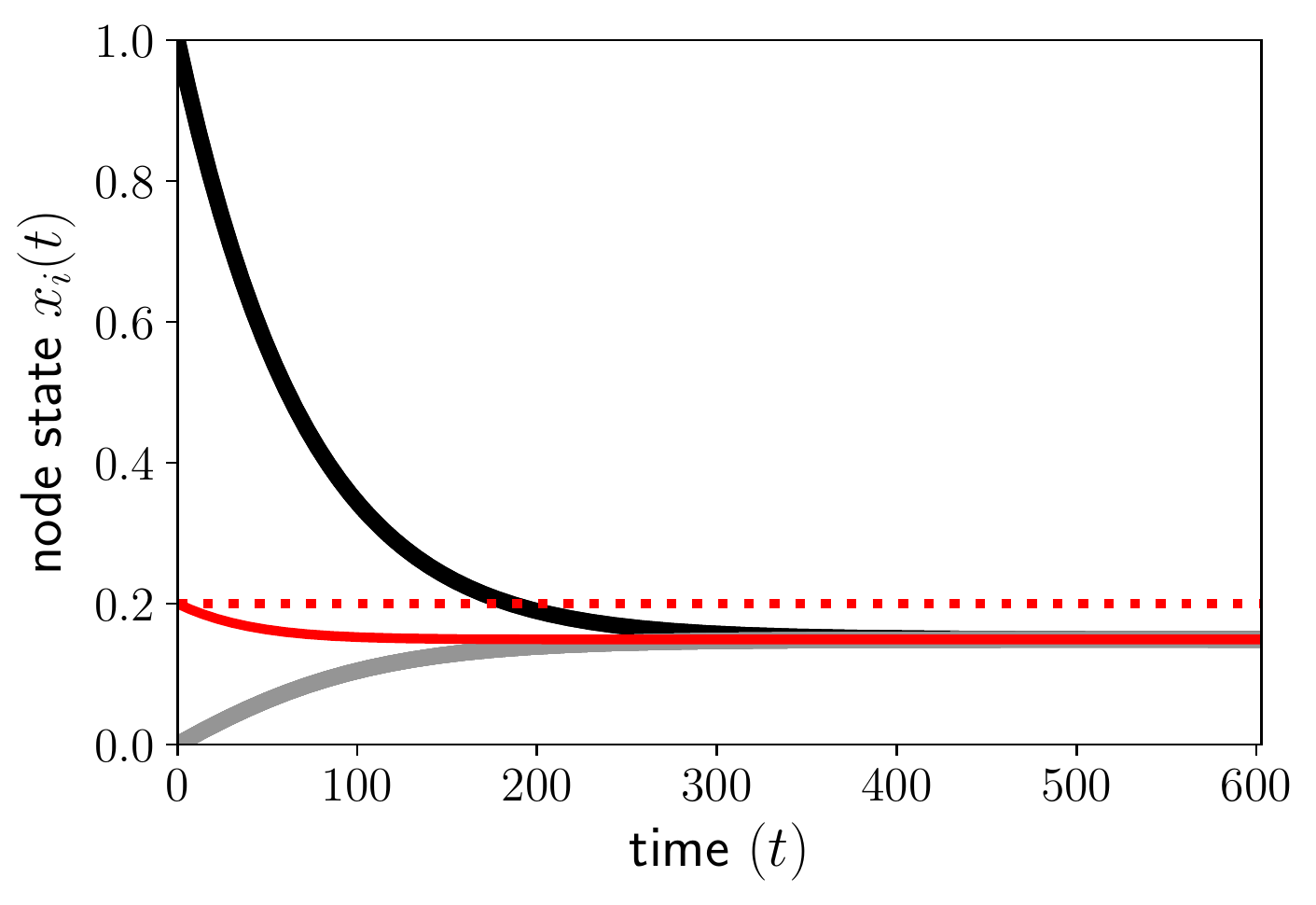}%
		}
		\hfill
		\subfloat[$\lambda=0$]{%
			\includegraphics[width=0.5\linewidth]{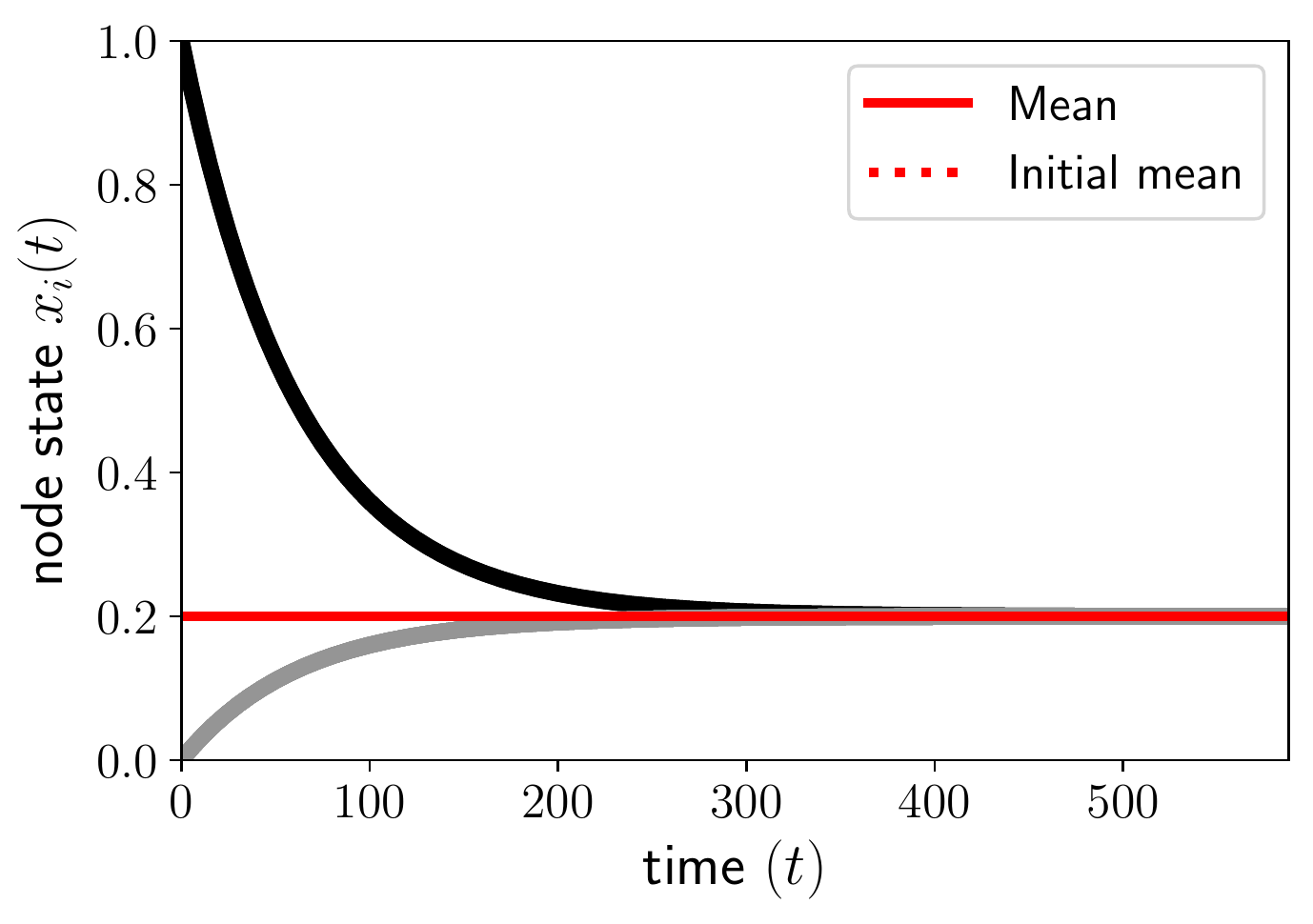}%
		}
		\caption{
			\textbf{3CM dynamics on a hypergraph with all-to-all connectivity}
			We display simulations for a modulation function $s(x)=\exp(\lambda x)$ for different values of $\lambda$, and an unbalanced initial condition in which 80\% of the nodes have opinion $0$ and 20\% have opinion $1$ ($\bar{x}(0)=0.2$)
			Dotted red lines indicate the initial value of the  average node state. 
			Black (grey) solid lines represent the evolution of the state of nodes whose initial configuration is one (or zero). 
			(a) The setting $\lambda<0$ will results in a final consensus value that is shifted away from the initial average.
			(b) In contrast, the average state is conserved for $\lambda=0$ no matter what the initial average was (as the dynamics is linear). 
		}
		\label{fig:simulations_meanfield_biased}
	\end{figure}
	
	\subsection{Shifts towards majority opinions on complete graphs}
	To exhibit how the non-linear reinforcing diffusion dynamics \eqref{eq:dynamics} can lead to a shift of the average state over time, we first study the dynamics on a structurally featureless `complete' hypergraph in which all triplet interactions are present.
	We split the nodes into two factions with binary opinions $x_i = 0$ or $x_j=1$, respectively.
    In this case, for initial distributions with average $\bar{x}(0)=0.5$, the Laplacian $L^\mathfrak{W}$ will correspond to a balanced graph for which in-degree equals out-degree for all nodes, and accordingly the final consensus value will be the average of the initial opinions, which is invariant in this case.
	In contrast, initial distributions with $\bar{x}(0)\neq 0.5$, necessarily lead to an unbalanced graph Laplacian $L^\mathfrak{W}$. 
	We remark that these conditions depend on the regular topology of the system and will not hold for systems with more general interaction structure.
	
	Note that if $\bar{x}\neq 0.5$ the initial groups must have different size, with one group being a relative majority or minority, respectively.
	Let us now consider a decreasing modulation function $s(x)=\exp(\lambda x)$ with $\lambda < 0$, such that the opinions of similar nodes reinforce each other.
    In this case, any deviation from an initial average $\bar{x}(0) =0.5$ grows in time, with a drift towards the majority. 
	This is shown in Figure \ref{fig:simulations_meanfield_biased}.
	In the context of opinion dynamics, this type of dynamics may be interpreted as a kind of peer-pressure, which causes the average opinion in the system to shift towards the initial majority opinion. 
	
	\subsection{Opinion dynamics in clustered systems}
\begin{figure}
		\centering
		\includegraphics[width=0.7\textwidth]{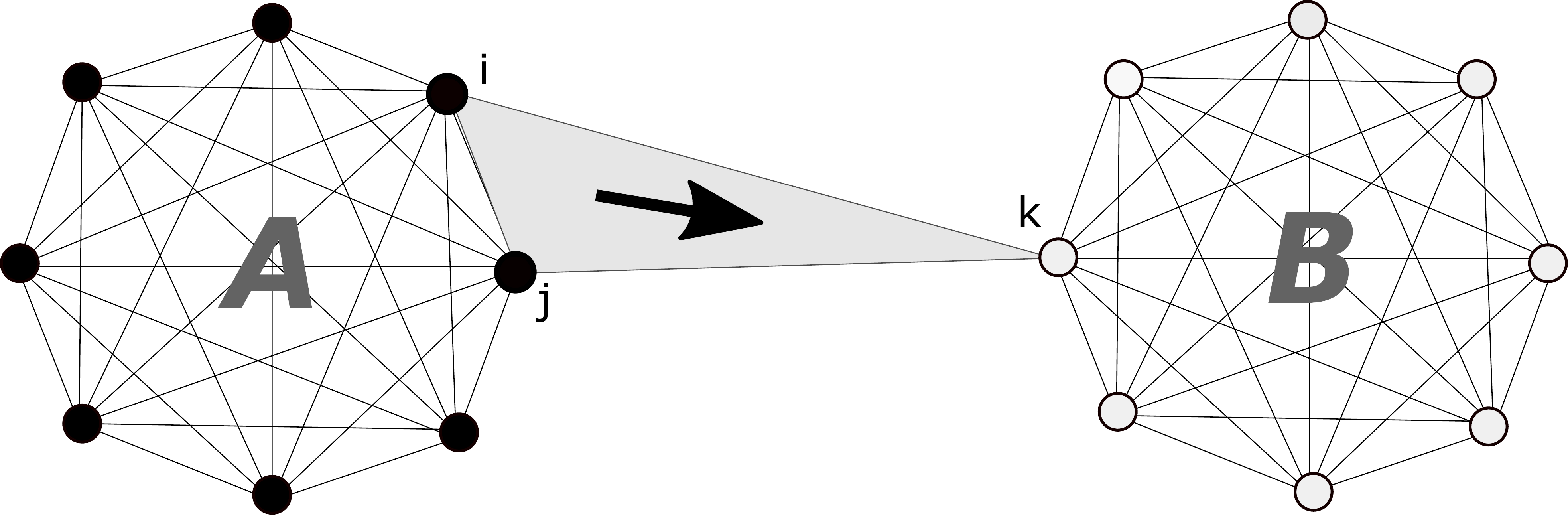}
		\caption{\textbf{Schematic: 3CM dynamics on a modular hypergraph.}
			We consider a binary initialisation of two clusters, above indicated as black and white node colours.
			For a  triplet interaction $(i,j,k)$ oriented towards cluster $B$ (top), the fact that there is a close consensus in cluster $A$ between nodes $i,j$ accelerates the rate of change of node $k$ in $B$. 
			In contrast, the influence on the nodes $i,j$ in $A$ is small, since the nodes in the pairs $(i,k)$ and $(j,k)$, belong to different clusters and are thus in relative disagreement, which decreases the effect of cluster $B$ on $A$.
		}
		\label{fig:binary_cluster}
	\end{figure}

	To gain some insight into the interplay between the system structure and our dynamics, we consider a system defined on simple modular hypergraph as displayed in Figure~\ref{fig:binary_cluster}. 
	Here the system consists of two fully connected equally sized clusters of $10$ nodes, i.e., $\mathcal{A}_{ijk} = 1$ for all triplets $(i,j,k)$ that are in the same cluster and $\mathcal{A}_{ijk}=0$, otherwise.
	In addition, the clusters are connected by $80$ randomly chosen triplet interactions, of which a fraction $p \in [0,1]$ is oriented towards cluster B, and the remainder towards cluster A. 
    Here we say a triplet interaction $(i,j,k)$ is oriented towards a specific cluster if two of its nodes are in the oppositve cluster (see Figure~\ref{fig:binary_cluster}).
    For instance, if nodes $i,j$ are in cluster $A$ and node $k$ is in cluster $B$, the triplet interaction is oriented towards cluster $B$ (recall, however, that all nodes of the triplet interact, i.e., there is no `directionality' in the coupling tensor $\mathcal A$).
	We consider an initially polarized state of opinions such that nodes in cluster A have initial state $x_A(0)=0$ and nodes in cluster B have initial  state $x_B(0)=1$. 
	
	In contrast to the fully connected system described in the previous section, here an initial average of $\bar{x}(0)=0.5$ does not guarantee that the average is invariant over time (i.e., the induced graph is not balanced).
	This is due to the reinforcing group effects which result in asymmetric interactions as encoded by the induced graph described by $\mathfrak{W}$.
	To see this, consider a triplet interaction oriented towards cluster B (see Figure \ref{fig:binary_cluster}.
	Since there is (local) consensus in cluster A, for monotonically decreasing $s(x)$ the influence of the nodes from A onto the node in B is increased, whereas the opposite influence is decreased.
	Following this reasoning, the relative influence of cluster A on B (and vice versa) thus depends on the relative number of triplet interactions oriented towards each cluster, which is determined by parameter $p$: for small $p$ most of the 'cross-cluster' triplet interactions will be oriented towards A, for large $p$ most interactions will be oriented towards B.

	\begin{figure}
		\centering
		\subfloat[Consensus value]{
			\includegraphics[width=0.5\linewidth]{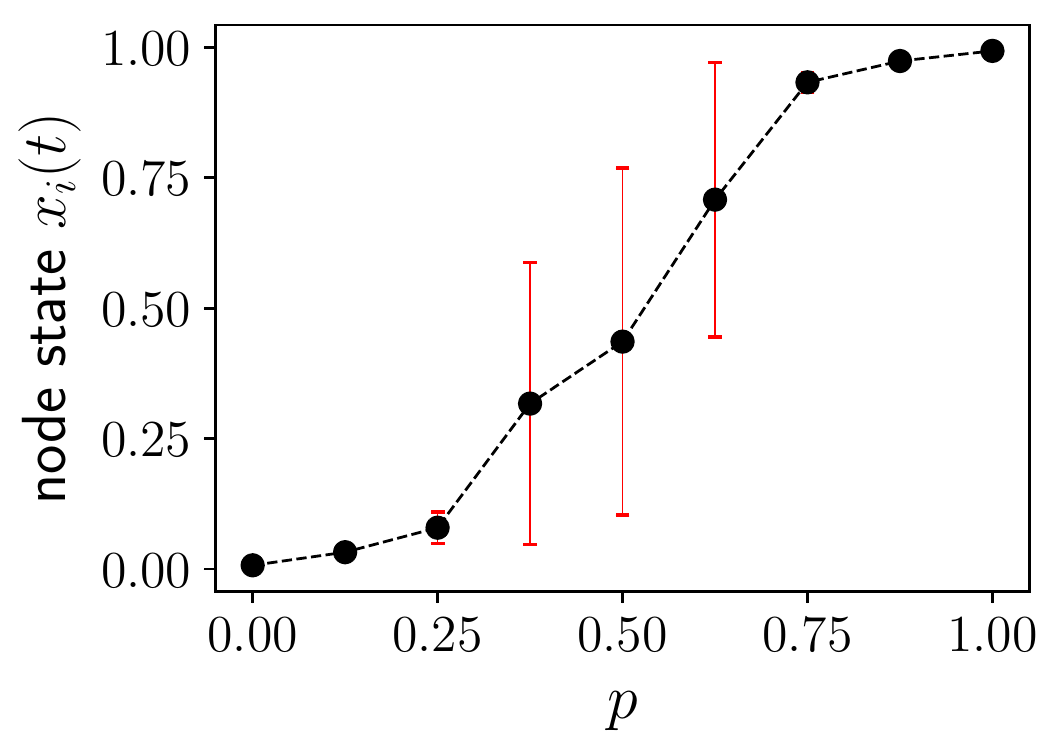}}
		\subfloat[Time until consensus]{	\includegraphics[width=0.5\linewidth]{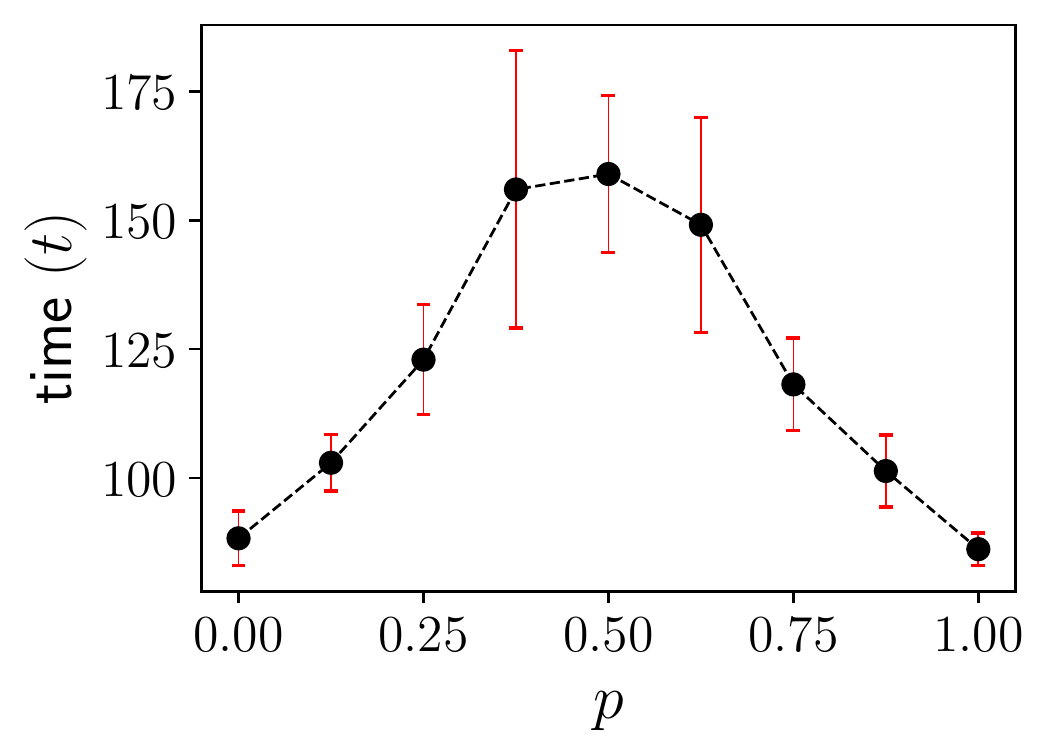}}
		\caption{\textbf{3CM dynamics on a modular hypergraph}
			Simulations of 3CM on two inter-connected clusters of 10 nodes, with  the modulation function  $s(x)=\exp(-100x)$.
			All results are averaged over 20 simulations, where the error bars denote one standard deviation.
			(a) Final consensus value as a function of the directionality parameter $p$. 
			As the fraction of triplet interactions oriented from cluster A to cluster B increases, so does the consensus value towards the initial state in cluster A.
			(b) The rate of convergence is significantly faster when the triplet interactions are mostly oriented towards one cluster, i.e., for extreme values of $p$. 
		}
		\label{fig:p_exp}
	\end{figure}
	
	In Figure \ref{fig:p_exp}, we show how the relative number of oriented triplet interactions measured by $p$ affect the final consensus value and the convergence towards consensus.
    For these results we averaged our 20 simulations with varying $p$.
	As seen in Figure~\ref{fig:p_exp}(a), we observe a shift in the final consensus value towards the initial value in cluster $A$ or in cluster $B$, depending on the percentage of triplet interactions oriented away from that cluster.
	The asymmetry also influences the rate of convergence towards consensus, as shown in Figure~\ref{fig:p_exp}(b), i.e., a relative increase in oriented triplets leads to a faster rate of convergence.
	Our simulations also reveal higher fluctuations in the asymptotic state for values close to $p=0.5$.
	This result indicates that the process is sensitive to small deviations from balance in the initial topology, which can lead to comparably large differences in the consensus value.
	
\begin{figure}
		\centering
		\subfloat[non-linear]{%
			\includegraphics[width=0.5\linewidth]{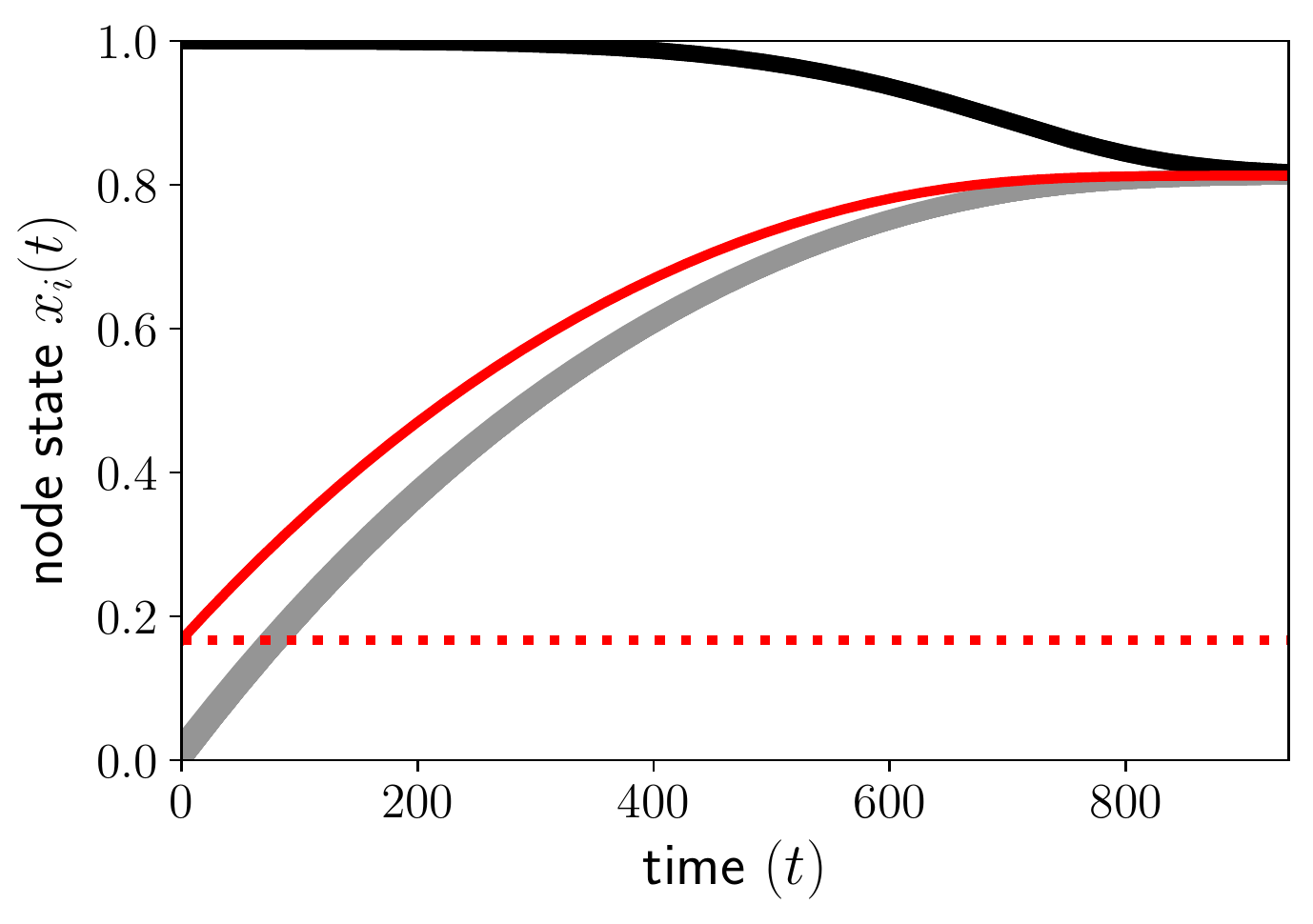}%
		}
		\hfill
		\subfloat[linear]{%
			\includegraphics[width=0.5\linewidth]{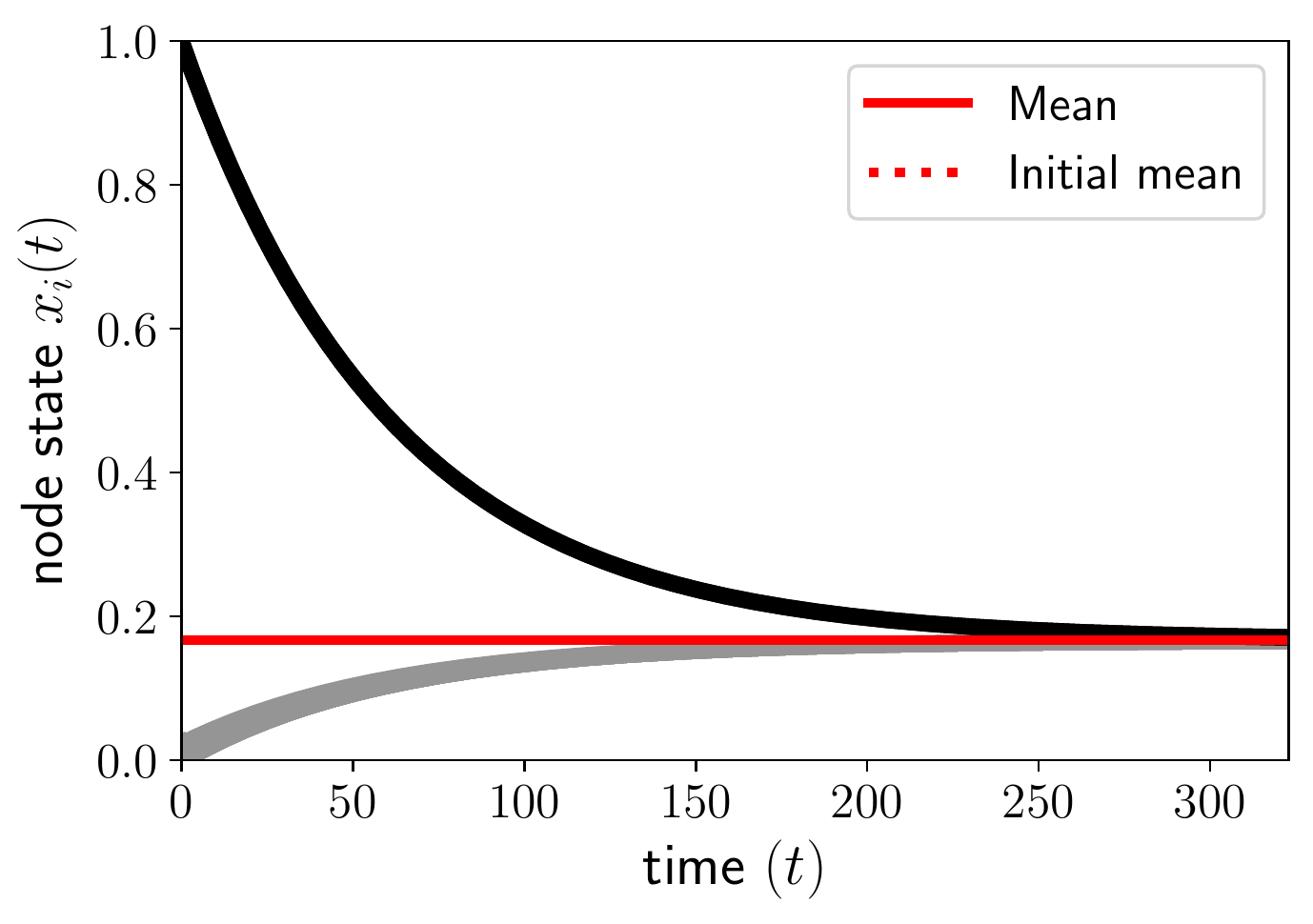}%
		}
		\caption{
			\textbf{Minority influence through reinforcing opininon.}
			We display simulations for a modulation function $s(x)=\exp(\lambda x)$ for $\lambda=-10$ and two biased clusters of different sizes, which are connected with a single triplet interaction oriented towards cluster A. 
            Cluster A comprises the majority of nodes (10 nodes) whereas cluster B consists only of 5 nodes.
            (a) While intuition may suggest a final consensus that is leaning towards the initial opinion $0$ of the majority cluster A, we observe the opposite behavior due to opinion reinforcing effect of the nonlinear coupling, which leads to an (effectively) oriented connection between B and A. 
            (b) If the dynamics are linear (right), the initial average opinion is conserved and therefore the majority opinion dominates the final consensus value.
		}
		\label{fig:minorities}
	\end{figure}

	Note that similar observations can also be made if one cluster forms a ``minority'' and is comparably smaller than the other cluster (the majority). 
	Indeed, depending on the relative number of triplets oriented towards the majority, the opinion of the minority cluster may have a much stronger influence on the final consensus value than the majority. 
    An example of this situation is shown in Figure~\ref{fig:minorities}.
    Here the opinion of the minority cluster A with 5 nodes `dominates' the opinion of the majority cluster B with 10 nodes for a non-linear reinforcing modulation function $s(x)$ (left).
	In the context of opinion dynamics, this type of behavior is akin to a ``minority influence'', in which small groups can dominate the formation of an opinion not because of their size, but due to their internal cohesion.
    In contrast, if we remove the effect of opinion reinforcement and consider a linear coupling the initial opinion of majority will have the strongest effect on the final consensus state (as expected from a distributed averaging).
	
	\subsection{Time-scale separations in clustered systems and multi-body interactions}
    Finally, we investigate interplay between the topology in a clustered hypergraph and our multi-body interaction dynamics for initial conditions that are not piecewise constant. 
    Specifically, we are interested in examining the different time scales in the dynamics induced by the clustered topology (as is also well known for pairwise interaction systems in the fixed \cite{Chow2016TimeSM} and time-varying case \cite{Martin2016TimeSMdynamic}).
    The different time-scales are here associated to a fast convergence of states inside clusters, followed by a slower convergence towards global consensus.

\begin{figure}
		\centering
		\begin{subfloat}{
				\includegraphics[width=0.45\linewidth]{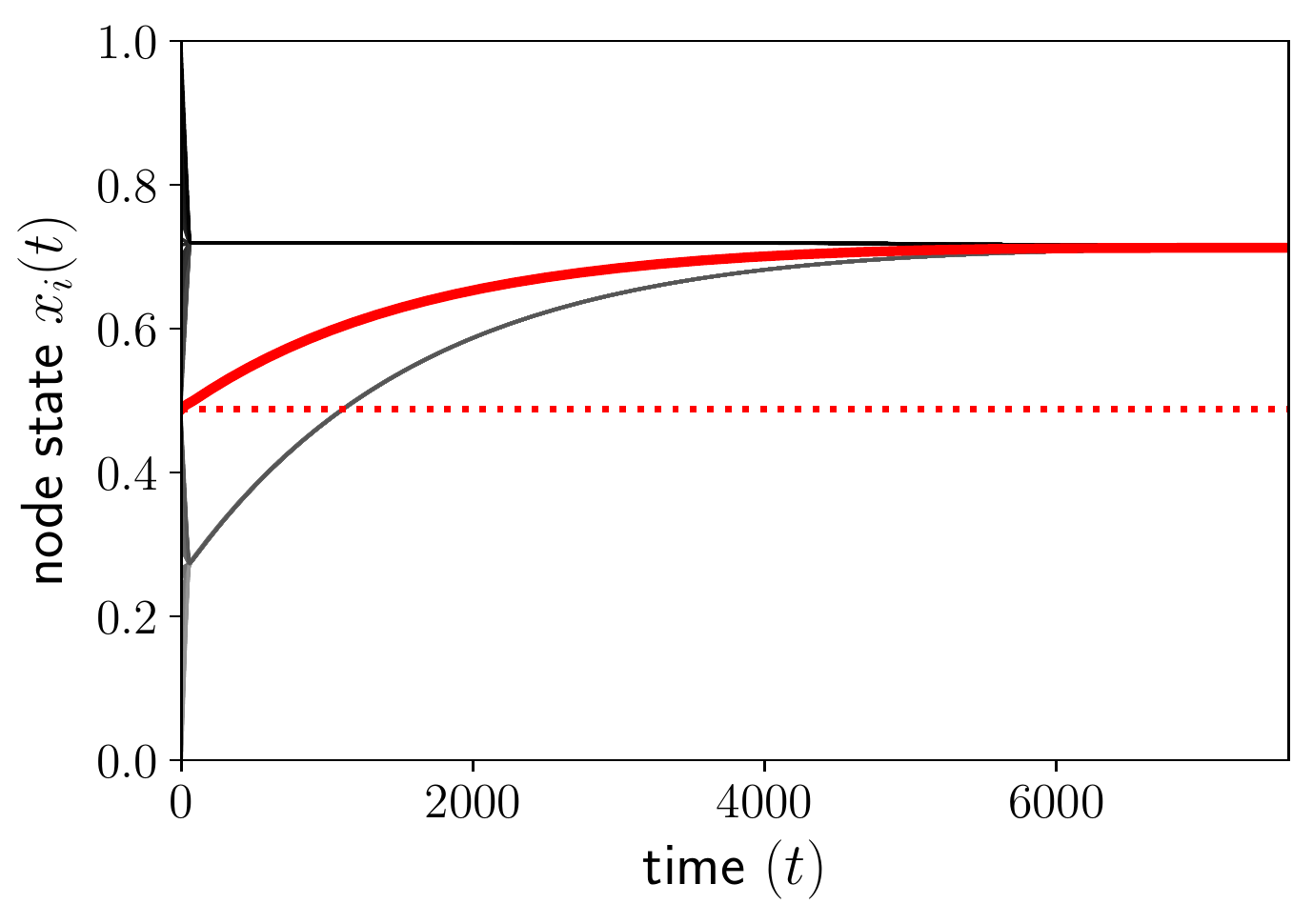}}
		\end{subfloat}
		\begin{subfloat}{
				\includegraphics[width=0.45\linewidth]{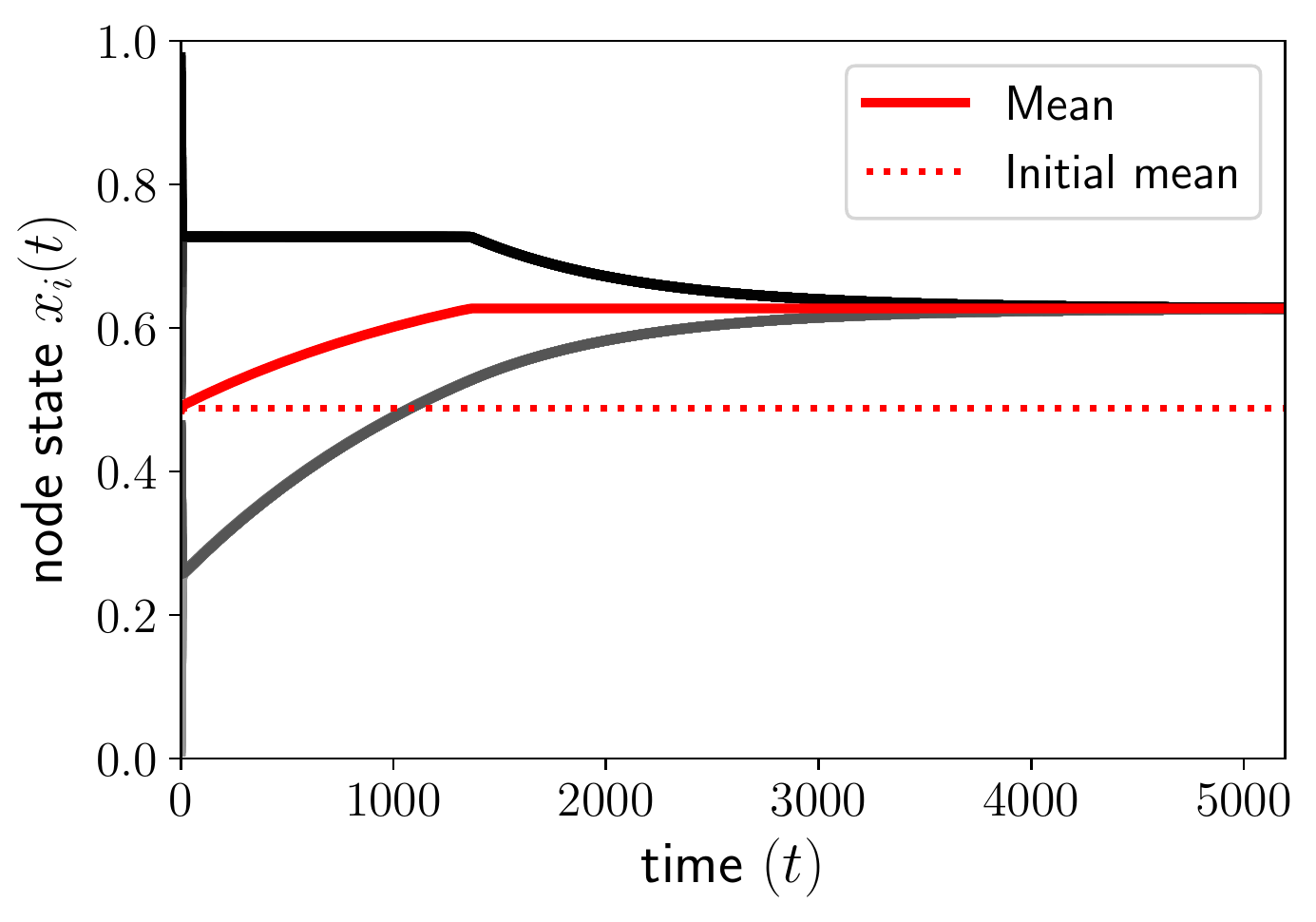}}
		\end{subfloat}
		\begin{subfloat}{
			\includegraphics[width=0.45\linewidth]{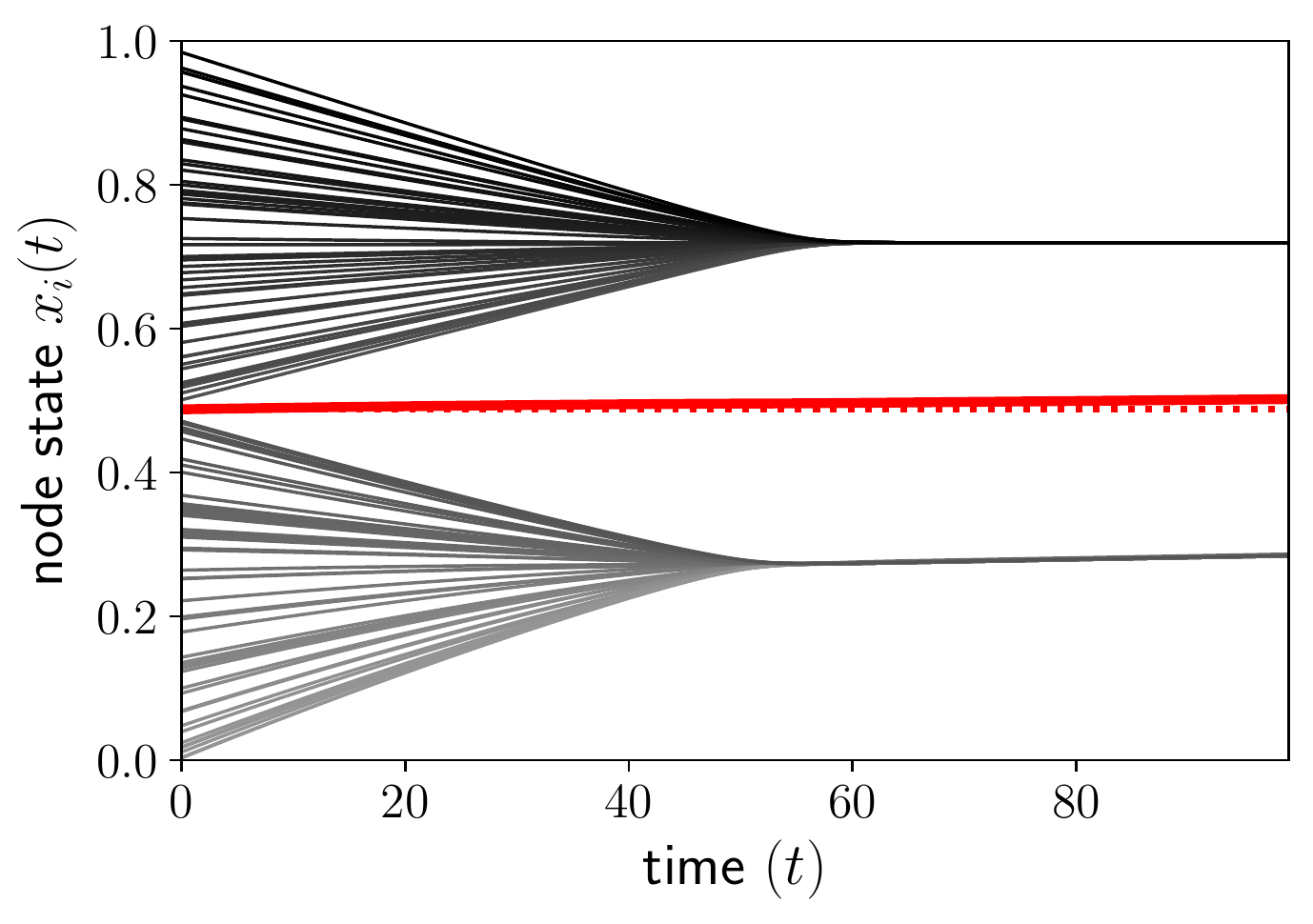}}
		\end{subfloat}
		\begin{subfloat}{
				\includegraphics[width=0.45\linewidth]{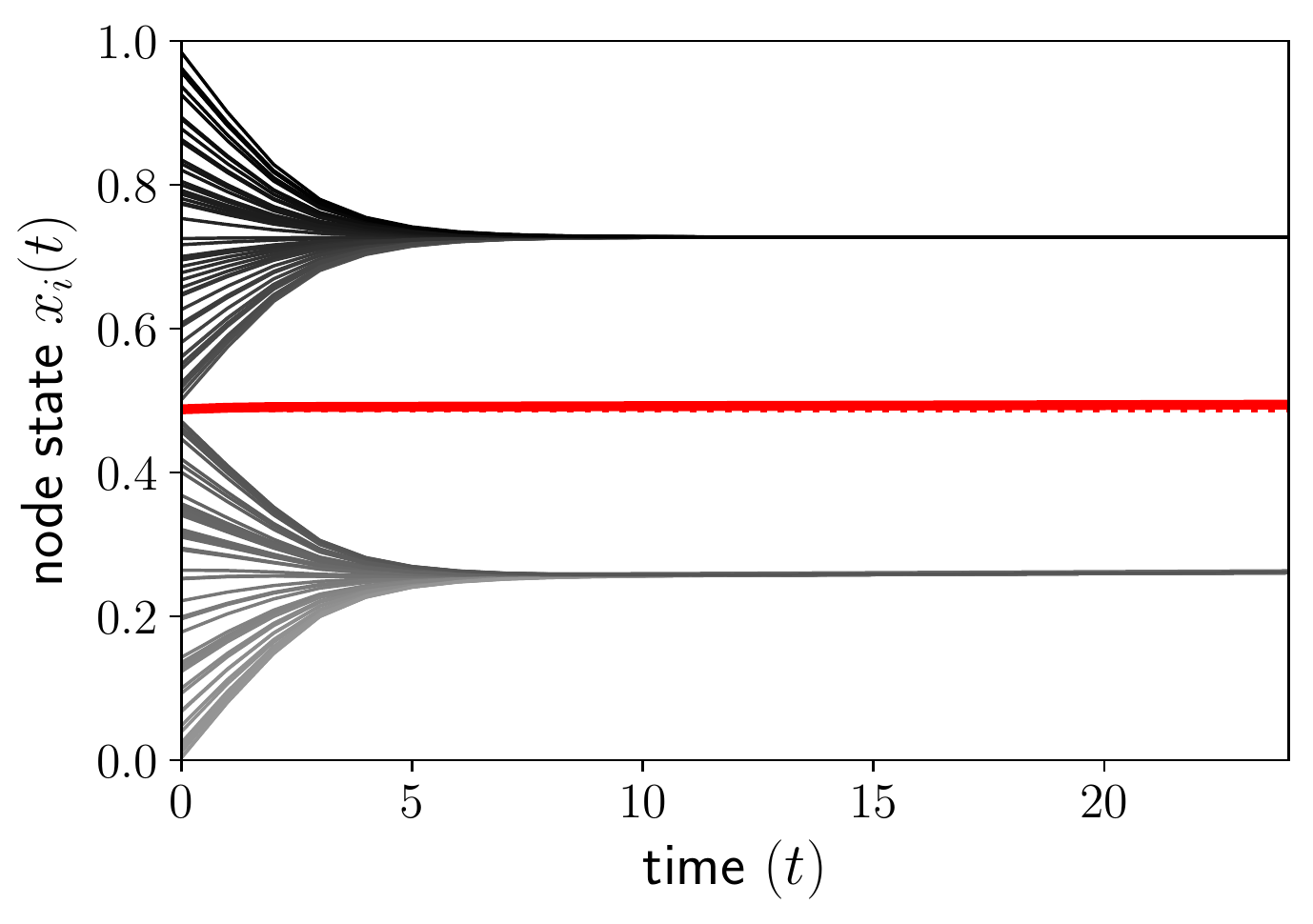}}
		\end{subfloat}
		\caption{
            \textbf{Time-scale separation in clustered systems and the influence of the modulation function $s(x)$.}
            We simulate the dynamics of two clusters $A$ and $B$ connected with 80 (random) triplet interactions all oriented towards nodes in $A$.
            The nodes states $x_i$ are initialised uniformly at random over two separate intervals $I_A, I_B$ with $I_A \cap I_B = \emptyset$, such that $x_i(0) \in I_A$ if $i$ is in cluster A and $x_i(0) \in I_B$ if $i$ is in cluster B.
            The left figures correspond to the exponential modulation function $s(x)=\exp(\lambda x)$ for $\lambda=-100$.
            The right figures correspond to a Heaviside modulation function with threshold  $\phi=0.2$.
            In both cases, we observe a timescale separation with a fast, symmetric dynamics inside the clusters, followed by a slow dynamics in which cluster $B$ exhibits a disproportionate influence compared to its size (both cluster have the same size).
            For the Heaviside function, the process becomes linear when the values in the two clusters are less separated than the Heaviside-threshold.
            The fast transient inter-cluster dynamics is shown in the bottom figures, with qualitatively similar results for both modulation functions.}
		\label{fig:timescale_sep}
	\end{figure}

    For simplicity, we consider here again the setting of a clustered hypergraph with 2 clusters described above for $p=1$ (cf. Figure~\ref{fig:binary_cluster} and Figure~\ref{fig:p_exp}).
    This time, however, the nodes in each cluster may have different states initially.
    For our experiments, we initialise nodes in different clusters uniformly at random over disjoint intervals, such that nodes of cluster $A$ have random initial states in the interval $I_A=[0,0.5]$ and nodes in cluster $B$ have random initial states in $I_B=[0.5,1]$ (see \Cref{fig:timescale_sep}).
	The initial cluster averages of the node states are thus far apart.

    Now two effects lead to a fast multi-body consensus inside each cluster.
	First, each of the clusters are internally fully-connected.
    Second, the inter-cluster-dynamics will generally have a weaker effect since the difference of the initial conditions means that $s(\|x_i - x_j\|)$ will be small if nodes $i$ and $j$ are in different clusters. 
    As a result, we first observe a fast dynamics within the clusters in which nodes approach the cluster-average state (\Cref{fig:timescale_sep}, bottom) and then a slower dynamics between the two clusters (\Cref{fig:timescale_sep}, top).

    However, the final outcome of this process critically depends on the modulation function $s(x)$.
    For $s(x)=\exp(\lambda x)$, with $\lambda=-100$, we observe an asymmetric shift towards cluster B for $p=1$, the final opinion as shown in \Cref{fig:timescale_sep} (left). 
    If we consider other modulation functions, however, the results can be different.
    For instance we may consider a coupling via a (shifted) heaviside function of the form:
	\begin{align}
	s(\|x_j-x_k\|)=H(\|x_j-x_k\|-\phi)=\begin{cases}0 & \mbox{if } \|x_j-x_k\|<\phi \\
	1 & \mbox{otherwise},
	\end{cases}
	\end{align}
	which switches between a zero interaction and linear diffusion when the difference of the neighbouring triangle nodes becomes smaller than a threshold $\phi \in (0,1)$.

    If we consider the Heaviside function with threshold $\phi = 0.2$ as the modulation function, the dynamics between the clusters behaves initially very similar to the exponential case (see \Cref{fig:timescale_sep}).
    This behavior continues until the two cluster means are less separated than $\phi$. 
    As shown in \Cref{fig:timescale_sep} (right), the dynamics between the clusters then become linear and therefore the average opinion remains constant from then onwards.

	We remark that the Heaviside function is not positive-definite, so that the above dynamics do not necessarily converge to a global consensus asymptotically.
    Indeed, consider a setting with three clusters that are connected by a triplet interaction with exactly one node in each cluster.
    If the initial states are separated more than the Heaviside constant, all inter-cluster interactions would be zero and therefore the system would at most converge to a decoupled `polarised' state with three independent opinions, one for each cluster.
	
	\section{Conclusion}
	We have explored a model for opinion formation with multi-body interactions, defined on hypergraphs, in order to identify the impact of reinforcing opinions on the dynamics. 
    We found that these non-linear multi-body interactions can cause dynamical phenomena such as shifts in the average opinion that would not appear in a corresponding pairwise system. 
    In situations with two connected, polarised groups the dynamics can lead to the opinion of one group clearly dominating the other. 
    These findings are important to better understand processes governed by reinforcing group effects in society. 
	
    In standard linear opinion formation models with two-body dynamical system such as the de-Groot model~\cite{DeGroot1974} or consensus dynamics~\cite{Tsitsiklis1984,olfati2007consensus}, it is known that the asymptotic behavior of the dynamics is dominated by the networks structure.
    For instance, the mixing time is determined by the spectral gap. 
    In contrast, in our model knowing the structure alone is not sufficient to understand the asymptotic behavior.
    Indeed, we have shown the initial node states can lead to an effectively oriented flow in the dynamics and thus lead to an opinion formation process that can be dominated, even by relatively small groups of nodes --- provided, they have a coherent opinion.
    In real-world setings, such an insight could be exploited to steer the opinion towards a desirable outcome 
    by modifying the balance in the dynamical system by seeding (i.e., changing the initial states) or eliminating components (i.e. changing the topology). 
    Future work will consider such issues in more detail. 
    A specific challenge in this context will be the derivation of simple, computable (heuristic) strategies that would enable for such a control, without having to assees all minute details of the hypergraph and initial condititions.
    For instance, deriving some form of higher-order generalization of network centralities, related to the dynamical properties of such multi-body interation systems, would be an interesting avenue to pursue.
    As a first attempt one may make use of the matrix represention $\mathfrak{W}$ of the 3CM system to derive appropriate centrality values.
    However, this would assume a fixed initialisation is known --- one thus would need to generalize this notion to identify the important actors and connections in the 3CM system.
	\bibliographystyle{splncs04}
	\bibliography{references}
\end{document}